\documentclass{ws-procs975x65}            
\begin{document}
\title{The Strong CP Problem and Discrete Symmetries}

\author{Martin~Spinrath}

\address{
Institut f\"ur Theoretische Teilchenphysik, Karlsruhe Institute of Technology,\\
Engesserstra\ss{}e 7, D-76131 Karlsruhe, Germany\\
E-mail: martin.spinrath@kit.edu\\
www.ttp.kit.edu}

\begin{abstract}
We discuss a possible solution to the strong CP problem
which is based on spontaneous CP violation and discrete symmetries.
At the same time we predict in a simple way the almost right-angled
quark unitarity triangle angle ($\alpha \simeq 90^\circ$) by making
the entries of the quark mass matrices either real or imaginary.
To prove the viability of our strategy we present a toy flavour model
for the quark sector.
\end{abstract}


\bodymatter

\section{Motivation}

It is fair to say that quantum chromodynamics (QCD) has emerged as the well-established
theory of strong interactions. However, there are still puzzles about the strong interactions.
One of them is the smallness of CP violation. Already in the 1970s it was realised that the
QCD Lagrangian can violate CP due to instanton effects \cite{Belavin:1975fg, 'tHooft:1976up}
which is described by the strong phase
\begin{equation} \label{eq:DefThetaBar}
 \bar \theta = \theta + \arg \det (M_u M_d) \;, 
\end{equation}
where $\theta$ is the coefficient of $\alpha_s/(8\pi) \tilde G_{\mu \nu} G^{\mu \nu}$,
$G_{\mu\nu}$ is the field strength tensor of QCD, $\tilde G_{\mu\nu}$ its dual,
and $\arg\det (M_u M_d)$ is the anomalous contribution from the quark masses.
While $\theta$ and $ \arg \det (M_u M_d)$ are transformed into each other via a
chiral transformation, the combination $\bar \theta$ stays invariant.
Experiments put stringent bounds on $\bar \theta \lesssim 10^{-11}$, see Ref.~\citenum{Beringer:1900zz, Burghoff:2011xk},
which is much smaller than the Jarlskog invariant, $J=\left(2.96^{+0.20}_{-0.16}\right)\times10^{-5}$, see Ref.~\citenum{Beringer:1900zz}.
Therefore, the essence of the strong CP problem is the question why the two contributions to
$\bar \theta$ sum up to such a small number.

There are three main ideas put forward to explain the smallness
of $\bar \theta$. The first and simplest solution is that one of the quarks is massless
\cite{'tHooft:1976up}. In this case the strong CP phase $\bar\theta$ is unphysical.
However, recent data strongly suggests that
all quarks are massive \cite{Beringer:1900zz}.

The second popular solution is the so-called axion
\cite{Peccei:1977hh} where $\bar \theta$ is promoted to
a dynamical degree of freedom which is set to small values
by a potential. This solution is very elegant but albeit
there have been extensive searches for axions there have
been no convincing experimental hints for their existence so far
\cite{Beringer:1900zz}.

The third approach solves the strong CP problem by breaking parity
(or CP) spontaneously. Then on the fundamental level  the term
$\alpha_s/(8\pi) \tilde G_{\mu \nu} G^{\mu \nu}$, which
violates parity as well as CP, is forbidden by either parity and/or CP,
which are assumed to be fundamental symmetries. For a short overview
and more references, see Ref.~\citenum{Antusch:2013rla}, where we
introduce the class of models discussed here.

As we discuss in the next section, where we outline our strategy,
our class of models is based on a sum rule for the phases in the 
CKM matrix \cite{Antusch:2009hq} suggesting a simple
structure for quark mass matrices with either real or purely imaginary
elements \cite{rightangledothers}. For an alternative class of textures
models see, for instance, Refs.~\citenum{Barr:1996wx,Barr:1997ph,Masiero:1998yi,
Glashow:2001yz,Chang:2003uq}.
As we will see our structure is realised in a simple manner in flavour models
based on discrete symmetries where the CP symmetry is spontaneously broken
using a method dubbed discrete vacuum alignment \cite{Antusch:2011sx}.
This method was used as well in various flavour models \cite{Meroni:2012ty, Antusch:2013kna}
which nevertheless usually put
a stronger focus on the lepton sector.

\section{The Strategy}
\label{sec:Strategy}

If CP is a fundamental symmetry of the Lagrangian, the strong CP phase
$\bar\theta$ vanishes on the fundamental level. However, in order to explain
CP violation in weak interactions, CP has to be broken spontaneously. And this
has to be done in a controlled way to keep the strong CP phase $\bar \theta$ at
least tiny enough to be in agreement with experimental data. 

In our class of models\cite{Antusch:2013rla} we have quark mass matrices with
$\arg \det (M_u M_d) = 0$
but still the value for the CKM phase is realistic.
Furthermore, we disfavour unnatural cancellations between the phases in the
up-type and the down-type quark sector. Hence, $\det M_u$ and $\det M_d$ should
be real (and positive) by itself already.

One possible choice is, for instance, that $M_u$ is completely
real and has a negligible 1-3 mixing (1-3 element), and that 
\begin{equation} \label{eq:MdStructure}
 M_d = \begin{pmatrix}
	0 & * & 0 \\
	* & \text{i}\,  * & * \\
	0 & 0 & *
       \end{pmatrix} \;,
\end{equation}
where $'*'$ are arbitrary but real entries. The only non-trivial complex phase appears
in the purely imaginary 2-2 element of $M_d$. Then the determinants of both mass
matrices are real.

This structure of the mass matrices can be realised from the spontaneous
breaking of CP and we indeed have a solution for the strong CP problem
as we will show in the following.
And furthermore this very simple structure can also correctly reproduce
the right quark unitarity triangle, as it was demonstrated in Ref.~\citenum{Antusch:2009hq}, 
since it satisfies the phase sum rule
\begin{equation}
\alpha \approx \delta_{12}^d - \delta_{12}^u \approx 90^{\circ}  \;,
\end{equation}
where $\alpha$ is the angle of the CKM unitarity triangle
measured to be close to $90^{\circ}$ \cite{Beringer:1900zz} and
$\delta_{12}^{d/u}$ are the phases of the complex 1-2 mixing angles
diagonalising the quark mass matrices (for the conventions used, see
Ref.~\citenum{Antusch:2009hq}). Now any model, which generates such a structure
could do the trick, and in the following we will discuss one possible example.

Suppose we have a (discrete, non-Abelian) family symmetry $G_F$ with triplet
representations (we use as an example $A_4$, but $S_4$, $T'$, $\Delta(27)$,
etc.\ would work equally well). See Ref.~\citenum{King:2013eh} for a recent review
on family symmetries. In our toy model we assume the right-handed down-type
quarks to transform as triplets under $G_F$ while all
other quarks are singlets. Then the rows of $M_d$ are proportional
to the vacuum expectation values (vevs) of family symmetry breaking
Higgs fields, so-called flavon fields, which are triplets under $G_F$ as well.
$M_u$ is generated by vevs of singlet flavon fields.

We introduce four flavon triplets with the following alignments in flavour space
\begin{equation}\label{eq:vevs}
 \langle \phi_1 \rangle \sim \begin{pmatrix} 1 \\ 0 \\ 0 \end{pmatrix} \;,\quad
 \langle \phi_2 \rangle \sim \begin{pmatrix} 0 \\ 1 \\ 0 \end{pmatrix} \;,\quad
 \langle \phi_3 \rangle \sim \begin{pmatrix} 0 \\ 0 \\ 1 \end{pmatrix} \;,\quad
 \langle \tilde \phi_2 \rangle \sim \text{i}\,  \begin{pmatrix} 0 \\ 1 \\ 0 \end{pmatrix} \;,\quad
\end{equation}
where we have explicitly shown the phases and which
can be achieved by standard vacuum alignment techniques.
Note that only $\tilde \phi_2$ has a complex (imaginary) vev.

To fix the phases of these vevs we use the method described in Ref.~\citenum{Antusch:2011sx},
which we want to sketch here for a singlet flavon field $\xi$. Suppose $\xi$ is charged
under a discrete $Z_n$ symmetry and apart from that neutral then we can write down
a superpotential for $\xi$
\begin{equation}
\mathcal{W} = P \left( \frac{\xi^n}{\Lambda^{n-2}} \mp M^2 \right) \;,
\label{eq:W-strucutre}
\end{equation}
where $P$ is a total singlet and $M$ and $\Lambda$ mass parameters.
We have dropped couplings for brevity and since we assume
fundamental CP symmetry the couplings and the mass parameters
are real.\footnote{Note that we use the generalised CP transformation, which is trivial with
respect to $A_4$. It agrees with the ordinary CP transformation for real representations of
$A_4$. See Refs.~\citenum{Feruglio:2012cw,Holthausen:2012dk} for a recent discussion
of generalised CP in the context of non-Abelian discrete symmetries.}
For the scalar potential for $\xi$ we find
\begin{equation} \label{eq:flavonpotentialZn}
V = |F_P|^2 =  \left| \frac{\xi^n}{\Lambda^{n-2}} \mp M^2 \right|^2 .
\end{equation}
and since $|F_P| \stackrel{!}{=} 0$  the vev of $\xi$ has to satisfy
\begin{equation}
\langle \xi^n \rangle = \pm \,\Lambda^{n-2}  M^2\;.
\end{equation}
This means
\begin{equation}\label{eq:phaseswithZn}
\arg(\langle \xi \rangle) =   \left\{ \begin{array}{ll}
\frac{2 \pi}{n}q \;,\quad q = 1, \dots , n & \mbox{\vphantom{$\frac{f}{f}$} for ``$-$'' in Eq.~(\ref{eq:flavonpotentialZn}),}\\
\frac{2 \pi}{n} q +\frac{\pi}{n} \;,\quad q = 1, \dots , n & \mbox{\vphantom{$\frac{f}{f}$} for ``$+$'' in Eq.~(\ref{eq:flavonpotentialZn}).}
\end{array}
\right.
\end{equation} 
Here the phases of the vevs do not depend on potential parameters, a situation
which has been dubbed 'calculable phases' in the literature\cite{Branco:1983tn}.
In Ref.~\citenum{Holthausen:2012dk} this was understood as the result of an
accidental CP symmetry of the potential.

Due to the stringent constraints on $\bar \theta$, special care needs to be taken
with possible corrections to this parameter. The most important corrections are:
\begin{enumerate}
\item Higher dimensional operators in the superpotential that could spoil the structure of the mass matrices and hence generate a non-vanishing $\arg \det (M_u M_d)$.
\item Corrections which are induced from the soft SUSY breaking terms.
\end{enumerate}
Here we are only going to touch the first point. For the second point
we refer to the discussion in Ref.~\citenum{Antusch:2013rla}.

\section{The Model}
\label{sec:Model}

In this section we briefly sketch the toy model presented in Ref.~\citenum{Antusch:2013rla}
which serves as a proof that the strategy outlined before can be realized in an explicit model
controlling higher dimensional operators in the superpotential.

As gauge symmetry we stick to the Standard Model gauge group and impose CP to
be a fundamental symmetry.
We choose here as non-Abelian discrete family symmetry $A_4$ which is frequently used
in flavour model building, since it allows to readily realise the observed large lepton mixing
(which we will not consider here) and since it is the smallest discrete group with triplet
representations. To avoid unwanted operators and to implement the discrete vacuum
alignment mechanism we have additionally the shaping symmetry $Z_4^5 \times Z_2 \times U(1)_R$.
The family symmetry is broken by the $\phi_i$, $i = 1,$ $2,$ $3$, and $\tilde{\phi}_2$
which are triplets under $A_4$, cf.~Eq.~(\ref{eq:vevs}). Additionally there are five singlet
flavons $\xi_i$, $ i = u,c,t,d,s$, which all receive real vevs.

To arrange for the flavon vev configuration to be dynamically realised along the lines outlined
in Sec.~\ref{sec:Strategy}, additional symmetries and fields have to be introduced.
This discussion is somewhat lengthy and technical such that we will skip the detailed discussion
of this nevertheless important ingredient. The interested reader can find the full superpotential
to align the flavon vevs in Ref.~\citenum{Antusch:2013rla}.

Instead we want to discuss in somewhat more detail the couplings of the flavons to the matter
sector and the corrections from higher-dimensional effective operators.
After symmetry breaking, the mass matrices will be generated by the superpotential
(remember that the right-handed down-type quarks form $A_4$ triplets while all other
matter fields are $A_4$ singlets)
\begin{align}
  \mathcal{W}_d &= Q_1 \bar d H_d \frac{\phi_2 \xi_d}{\Lambda^2} + Q_2 \bar d H_d \frac{\phi_1 \xi_d + \tilde \phi_2 \xi_s + \phi_3 \xi_t}{\Lambda^2}  +  Q_3 \bar d H_d \frac{\phi_3}{\Lambda} \;,\\
  \mathcal{W}_u &=  Q_1 \bar u_1 H_u \frac{\xi_u^2}{\Lambda^2}  +  Q_1 \bar u_2  H_u \frac{\xi_u \xi_c}{\Lambda^2}  +  Q_2 \bar u_2 H_u \frac{\xi_c}{\Lambda}   + (Q_2 \bar u_3 + Q_3 \bar u_2) H_u \frac{\xi_t}{\Lambda} + Q_3 \bar u_3 H_u\;,
\end{align}
which results from integrating out the heavy messenger fields and where
we dropped couplings for the sake of brevity. Trivial $A_4$ contractions are
not explicitly shown\footnote{The only non-trivial contraction is between $\bar d$
and the $\phi_i$, which form a singlet contracted by the $SO(3)$-type inner product
'$\cdot$'.}
and $\Lambda$ denotes a generic messenger scale which is larger than the family
symmetry breaking scale $M_F$.

Replacing Higgs and flavon fields with their respective vevs we find the
following quark mass matrices
\begin{align}
 M_d = \begin{pmatrix}
	0 & b_d & 0 \\
	b'_d & \text{i}\,  c_d & d_d \\
	0 & 0 & e_d
       \end{pmatrix}
\quad\quad\text{and}\quad\quad
 M_u = \begin{pmatrix}
	a_u & b_u & 0 \\
	0 & c_u & d_u \\
	0 & d'_u & e_u
       \end{pmatrix} \;.
\end{align}
where we use the left-right convention
$-\mathcal{L} = \overline{u^i_L} (M_u)_{ij} u^j_R + \overline{d^i_L}
(M_d)_{ij} d^j_R + \text{ H.c.}$. Note that due to the fundamental CP
symmetry and its peculiar breaking pattern, all entries are real apart from the
2-2 element of $M_d$. 
As discussed before in the strategy section, it predicts the right quark
unitarity triangle\cite{Antusch:2009hq}
in terms of a phase sum rule
\begin{equation}
\alpha \approx \delta_{12}^d - \delta_{12}^u \approx 90^\circ  \;, 
\end{equation}
where the angle $\alpha$ of the CKM unitarity triangle
is close to $90^\circ$ \cite{Beringer:1900zz}.

In this toy model, we concentrate on the explanation of CP violation in
strong and weak interactions. Therefore, we are content with the prediction
of the smallness of the strong CP phase and the correct CP phase in the CKM
matrix. We are able to fit all masses and mixing angles, cf.~Ref.~\citenum{Antusch:2009hq}.
A more realistic model should obviously aim at predicting the masses and mixing
angles as well, which happens quite naturally in a GUT context, for instance. In fact,
a similar texture has been obtained in a GUT based model in Ref.~\citenum{Antusch:2013kna},
which could solve the strong CP problem as well. 

We sketch now the UV completion of our toy model which justifies
completely the effective operators we have given before. We will
furthermore discuss all higher-dimensional operators which give
corrections to the mass matrices and to the flavon alignment.
They will not alter the structure of the mass matrices and hence
our conclusions remain unchanged.

\begin{figure}
\centering
\includegraphics[width=0.76\textwidth]{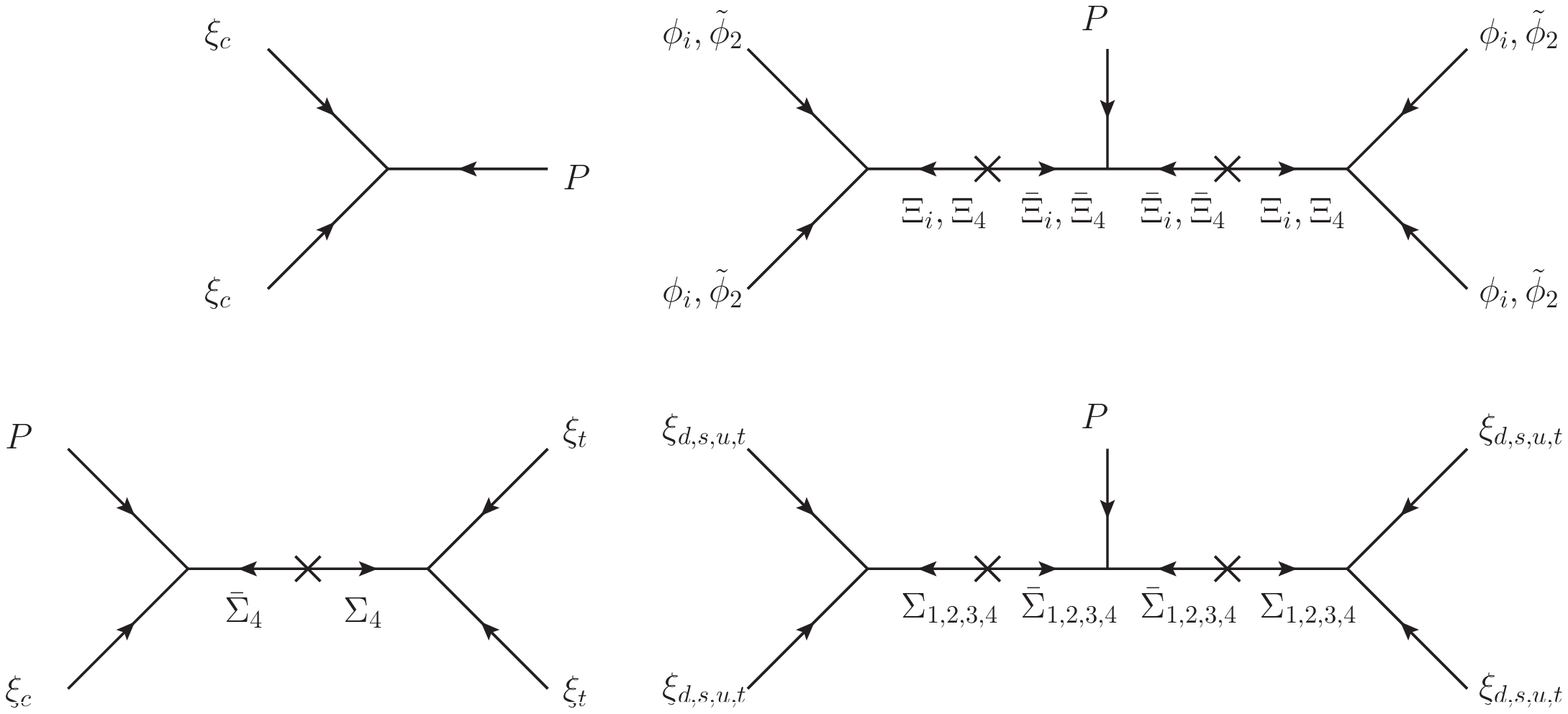}
\includegraphics[width=0.76\textwidth]{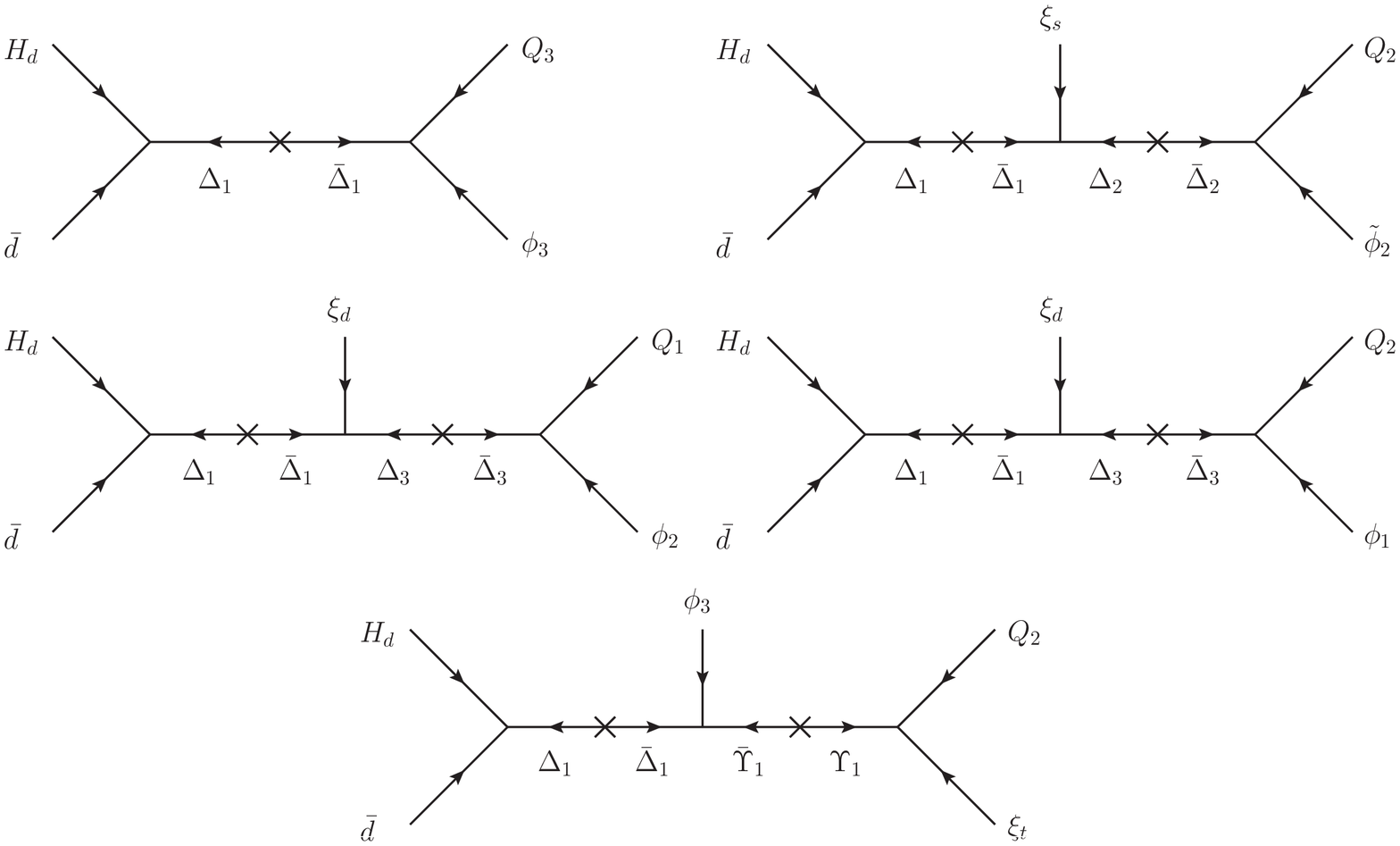}
\includegraphics[width=0.76\textwidth]{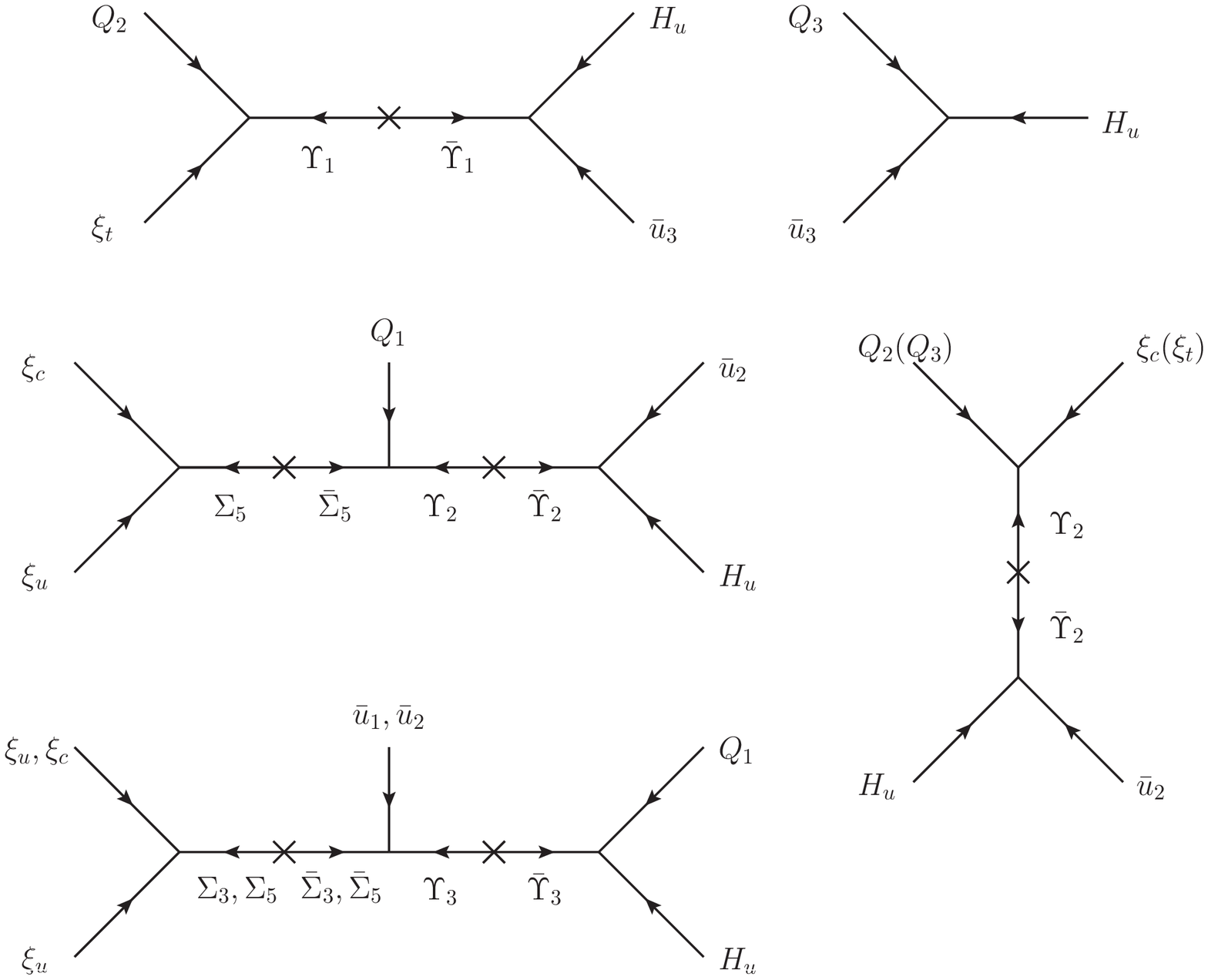}
\caption{
The supergraphs before integrating out the messengers in our model.
For the flavon sector only the diagrams are shown which fix the phases of the flavon vevs.
For more details, see Ref.~\citenum{Antusch:2013rla}.
\label{Fig:Messenger} }
\end{figure}

We will not go through the details of the full renormalisable superpotential here,
which can be found in Ref.~\citenum{Antusch:2013rla}. The relevant point is that
the symmetries and the chosen field content allow only for certain higher-dimensional
operators depicted by their respective supergraphs in Fig.~\ref{Fig:Messenger}.
After the heavy messenger fields are integrated out we end up first of all with the
leading operators which we needed to get the right flavon alignment and the right
quark mass matrices.

Beyond those operators we did not find  any higher-dimensional
operators produced at tree-level that would contribute to the down-type
quark sector.
In contrast, for the up-type quarks there are some additional
operators allowed which give (real) corrections to the
entries of the up-type quark mass matrix:
\begin{align}\label{eq:Wu-corr}
 \mathcal{W}^{\text{corr}}_{\text{u}} & = Q_1 \bar u_1 H_u \left( \frac{\xi_c^2 \xi_u^2 + \phi_1^2 \phi_2^2}{\Lambda^4} + \frac{\xi_c \xi_t^2 \xi_u^2}{\Lambda^5} + \frac{\xi_t^4 \xi_u^2}{\Lambda^6} \right)
+ Q_1 \bar u_2 H_u \frac{\xi_t^2 \xi_u}{\Lambda^3} + Q_2 \bar u_2 H_u \frac{\xi_t^2}{\Lambda^2} \;.
\end{align} 
These corrections are subleading real corrections to
real entries of the Yukawa matrix and hence do not
alter the fact that $\bar \theta=0$.

Comparatively complicated are the additional effective operators
for the flavon alignment
\begin{align}
 \mathcal{W}^{\text{corr}}_{\text{flavon}} & = \frac{P}{\Lambda^4} ( \phi_1^2 \phi_2^2 \xi_u^2 + \xi_u^4 \xi_c^2 )
 + \frac{P}{\Lambda^5} \xi_c \xi_t^2 \xi_u^4 \nonumber\\
 &+ \frac{P}{\Lambda^6} ( \xi_t^4 \xi_u^4 + \xi_c^2 \xi_u^2 \phi_1^2 \phi_2^2 + \phi_1^4 \phi_2^4 + \xi_u^4 (\phi_1^4 + \phi_2^4)) \nonumber\\
 &
 + \frac{P}{\Lambda^7} \xi_c \xi_t^2 \xi_u^2 \phi_1^2 \phi_2^2
 + \frac{P}{\Lambda^8} \xi_u^2 (\xi_c^2 \xi_u^2 (\phi_1^4 + \phi_2^4) + \phi_1^2 \phi_2^2 (\xi_t^4 + \phi_1^4 + \phi_2^4)) \nonumber\\
 &
 + \frac{P}{\Lambda^9} \xi_c \xi_t^2 \xi_u^4 (\phi_1^4 + \phi_2^4)
 + \frac{P}{\Lambda^{10}} (\phi_1^4 + \phi_2^4) (\xi_t^4 \xi_u^4 + \xi_c^2 \xi_u^2 \phi_1^2 \phi_2^2 + \phi_1^4 \phi_2^4) \nonumber\\
 &
 + \frac{P}{\Lambda^{11}} \xi_c \xi_t^2 \xi_u^2 \phi_1^2 \phi_2^2 (\phi_1^4 + \phi_2^4)
 + \frac{P}{\Lambda^{12}} \xi_t^4 \xi_u^2 \phi_1^2 \phi_2^2 (\phi_1^4 + \phi_2^4)
\;.
\end{align}
Nevertheless, a close inspection reveals
that our alignment including the phases of the flavon vevs is not altered by these
additional operators which can be supported by symmetry arguments\cite{Antusch:2013rla}.

Finally, let us briefly comment on the effects anomalies might have on our results.
The gauge symmetries remain anomaly free (after adding the leptons), because we do not
add new chiral fermions, which are charged under the Standard Model gauge group. 
In addition, as we do not introduce non-trivial singlet representations of $A_4$,
the $A_4$ group is anomaly free, but some of the auxiliary $Z_n$ symmetries appear
to be anomalous.\footnote{For a general discussion of anomalies of discrete symmetry
groups the reader is referred to Refs.~\citenum{Araki:2007zza,Ishimori:2012zz}.}
However, since we do not specify here a complete model (including leptons, a SUSY
breaking sector etc.), we cannot make definite statements about anomalies but we assume
that the effects of anomalies are either cancelled in the complete theory or sufficiently
small.

\section{Relation to Other Models}

In this section we want to discuss briefly how our
class of models is related to other models explaining
the smallness of the strong CP phase by a spontaneous
breaking of CP. We will especially focus on the Nelson-Barr
models of spontaneous CP violation\cite{Nelson:1983zb,
Barr:1984qx} being the first and most studied models. Although
there are certain similarities, our model,
for instance, does not fulfil the Barr criteria\cite{Barr:1984qx}.

We do not want to repeat the whole discussion. Instead, we just
give the mass matrices in both setups. In the Nelson-Barr setup
the mass matrix for the down-type quarks including
heavy vector-like quarks (what we call messenger fields) would read
\begin{equation}
 M_D \sim \begin{pmatrix} Y v_d & 0 \\ \langle \phi \rangle & M_\Upsilon \end{pmatrix} \;,
\end{equation}
where they assume $Y v_d$ and $M_\Upsilon$ to be real by
CP symmetry and only the vev of some symmetry breaking fields
which governs the couplings of the light to the heavy fields
induces CP violation. In such a setup one could get weak
CP violation while $\bar \theta \sim \arg \det M_D$ still
vanishes.

In our toy model we can explicitly write down the corresponding
mass matrix
\begin{equation}
 M_D \sim \begin{pmatrix}
      0 & 0 & 0 & \langle \phi_2 \rangle^T & 0 & 0 \\
      0 & 0 & \langle \tilde \phi_2 \rangle^T & \langle \phi_1 \rangle^T & \langle \xi_t \rangle & \langle \xi_c \rangle \\
      0 & \langle \phi_3 \rangle^T & 0 & 0 & 0 & \langle \xi_t \rangle \\
      \langle H_d \rangle & M_{\Delta_1} & 0 & 0 & 0 & 0 \\
      0 & \langle \xi_s \rangle & M_{\Delta_2} & 0 & 0 & 0 \\
      0 & \langle \xi_d \rangle & 0 & M_{\Delta_3} & 0 & 0 \\
      0 & \langle \phi_3 \rangle^T & 0 & 0 & M_{\Upsilon_1} & \langle \xi_t \rangle \\
      0 & 0 & 0 & 0 & 0 & M_{\Upsilon_2} \\
     \end{pmatrix} \;,
\end{equation}
which has the determinant
\begin{equation}
 \det M_D \sim \langle H_d \rangle^3 M_{\Delta_2}^3 M_{\Delta_3}^3 M_{\Upsilon_1} M_{\Upsilon_2} \langle \xi_d^2 \rangle \langle \phi_1 \rangle \langle \phi_2 \rangle \langle \phi_3 \rangle \;,
\end{equation}
which is real because $\langle {\tilde \phi}_2 \rangle$ does not appear.
This is only due to our alignment.

\section{Summary and Conclusions}
\label{sec:conclusions}

In this proceedings we have discussed a recently proposed
novel approach to solve the strong CP problem in the context
of spontaneous CP violation without the need for an axion.
We assume CP to be a fundamental symmetry of nature and use
discrete, Abelian and non-Abelian (family) symmetries to
break it in such a way that the anomalous contribution to
the CP violating QCD parameter $\bar \theta$ from the quark
mass matrices vanishes at tree-level. Simultaneously
the CKM phase is predicted to have its observed large
value in a simple and transparent way.

An essential ingredient of this approach is that the
phases of the symmetry breaking vevs are fixed to
certain discrete values with either being real or
purely imaginary in the simplest possible setup
which is governed in our example by the discrete
vacuum alignment method\cite{Antusch:2011sx}.
Nevertheless, other models reproducing the texture from
eq.~\eqref{eq:MdStructure} could do the same trick.

Our toy model is supersymmetric, which helps to
fix the flavon vev phases and forbids via
the non-renormalisation theorem the appearance of
new, unwanted operators in the superpotential
from loop corrections, which could spoil our solution
for the strong CP problem.
Furthermore, the model is  based on the family
symmetry $A_4$ with an $U(1)_R$ symmetry and the shaping
symmetry $Z_2 \times Z_4^5$ forbidding unwanted operators
and providing a mechanism to fix the phases of
the flavon vevs via the discrete vacuum alignment method.
We discussed an UV completion of the model in that sense
that we give a list of heavy messenger fields which generate
the desired effective operators after being integrated out.
This enables us to show explicitly that our solution for the
strong CP problem is not affected by higher order corrections
(ignoring non-perturbative and SUSY breaking effects).

Finally, we discussed the relation between our novel class
of models to the well known Nelson-Barr models\cite{Nelson:1983zb, Barr:1984qx}.
In the Nelson-Barr models direct couplings between the light sector and the heavy
sector are partially forbidden in such a way that the total mass
matrix exhibits a special block structure. This is different in our
class of models, where all light fields can couple to all heavy
messenger fields in principle. The determinant of the total mass
matrix in their case is real due to the mentioned block structure,
while in our case it is real due to our vacuum alignment (including
phases).

The class of models presented here casts new light on an old
problem, the strong CP problem. There have been several previous
attempts to solve it in terms of spontaneous CP violation in
combination with flavour symmetries but our strategy differs
significantly from these previous approaches.
Most notably, we simultaneously have large CP violation in the
CKM matrix with a right-angled unitarity triangle in a simple way,
without any contribution to $\bar \theta$ from the quark mass matrices.
Furthermore, the techniques to handle the symmetry breaking of
discrete non-Abelian family symmetries, like in our example model
$A_4$, was first developed in the context of the large leptonic
mixing angles and finds here an unexpected new application. Also
the method to fix the flavon vev phases was developed to give a
dynamical explanation for the phase sum rule but was then in
succeeding papers used in the lepton sector as well.

\end{document}